# Reconfigurable continuous-zoom metalens in visible

Yuan Cui[1], Guoxing Zheng[1,2], Ming Chen[3], Yilun Zhang[3], Yan Yang[4], Jin Tao[2], Taotao He[1], Zile Li[1,2,*]

[1] *Electronic Information School, Wuhan University, Wuhan, 430072, China*
[2] *NOEIC, State Key Laboratory of Optical Communication Technologies and Networks, Wuhan Research Institute of Posts and Telecommunications, Wuhan, 430074, China*
[3] *Photonics Research Centre, Guilin University of Electronic Technology, Guilin 541004, China*
[4] *Integrated Circuit Advanced Process Center, Institute of Microelectronics of Chinese Academy of Sciences, Beijing 100029, China*
*Corresponding authors: lizile@whu.edu.cn*

**Abstract:** Design of a conventional zoom lens is always challengeable because it requires not only sophisticated optical design strategy, but also complex and precise mechanical structures for lens adjustment. In this paper, we propose a continuous zoom lens consisting of two chiral geometric metasurfaces with dielectric nanobrick arrays sitting on a transparent substrate. The metalens can continuously vary the focal length by rotating either of the two metasurfaces along its optical axis without changing any other conditions. More importantly, because of the polarization dependence of the geometric metasurface, the positive and negative polarities are interchangeable in one identical metalens only by changing the handedness of the incident circularly polarized light, which can generate varyingfocal lengths ranging from −∞ to +∞ in principle. On account of its advantages of compactness, flexibility and easiness in design, the proposed zoom metalens can provide new perspectives for the development of continuous-zoom optical system and it can find applications in fields which require ultracompact and continuous-zoom imaging and reconfigurable beam wavefront steering.

## 1. Introduction

By virtue of the capability of manipulating the electromagnetic fields, metasurfaces [1-6] are widely used to develop many new optical devices and optimize traditional optical devices. In particular, geometric metasurfaces (GEMSs) get extensive attraction due to its precise phase control ability. The beauty of this approach lies in the linear dependence of phase delay $\varphi$ on the orientation angle $\phi$ of each nanostructure, i.e., $\varphi = \pm 2\phi$ [7-9], the sign is determined by the polarization state of incident circularly polarized light. More importantly, the scattering amplitude remains unchanged since the geometric nanostructure remains consistent. Therefore, GEMSs offer new perspectives in designing complex phase-only optical elements, such as holograms [10-18] and metalenses [19, 20] etc. As one of the most important optical functionalities, continuous-zoom has been widely valued by researchers, which is currently combined with metasurfaces to present a lot of designs, for example, zooming by stretching a flexible substrate [21], changing the focal length by lateral actuation [22] or a tunable metalens using microelectromechanical systems (MEMS), [23] etc. However, these zoom metalenses mentioned above have not done much work in terms of zoom range, and it is still not an easy task to tune focal length over a wide range for a metalens.

Recently, researchers have proposed the Moiré effect [24, 25] combined with diffractive optical elements (DOEs) to realize zooming by relatively rotating two DOEs [26, 27]. In this paper, we further verify that the Moiré effect can be realized by the combination of two chiral GEMSs and they can be used to form a continuous-zoom metalens. More interestingly, due to the polarization dependence [28-33] of a GEMS, the zoom range of the compound metalens can further be extended to −∞ ~ +∞, which indicates that we can shape a beam of incident light into a convergent, divergent or plane wave with any spherical radius in principle. Considering its advantages of ultracompactness, continuous zooming and easiness for adjustment and fabrication, the proposed continuous-zoom metalens can provide new perspectives and a practical method for wide-range and reconfigurable beam wavefront steering in an ultra-simple way.

## 2. Principle

The working principle of a continuous-zoom metalens consisting of two GEMSs is shown in Fig. 1(a). By rotating two GEMSs relatively, the focal length of the compound metalens varies continuously and the incident circularly polarized (CP) light can be focused on the image plane of the metalens. To form the two GEMSs, we use transmissive metasurfaces constructed by dielectric silicon nanobrick arrays sitting on a silicon dioxide substrate. The nanobrick arrays have the same dimension with length $L$, width $W$, height $H$, and unit-cell size $C$ as shown in Fig. 1(b). Each nanobrick acts as a half-wave plate [13, 34, 35] and it can transform the incident CP light into the opposite polarization state accompanied by a geometric phase delay $\varphi$ ($\varphi = \pm 2\phi$) that is demonstrated in the *Introduction*. Since the nanobrick's orientation angle can be designed in [0, $\pi$], the geometric phase is continuously and accurately controlled in [0, 2$\pi$] by assigning different orientation angles, which will bring great freedoms for phase-only device design such as lenses.

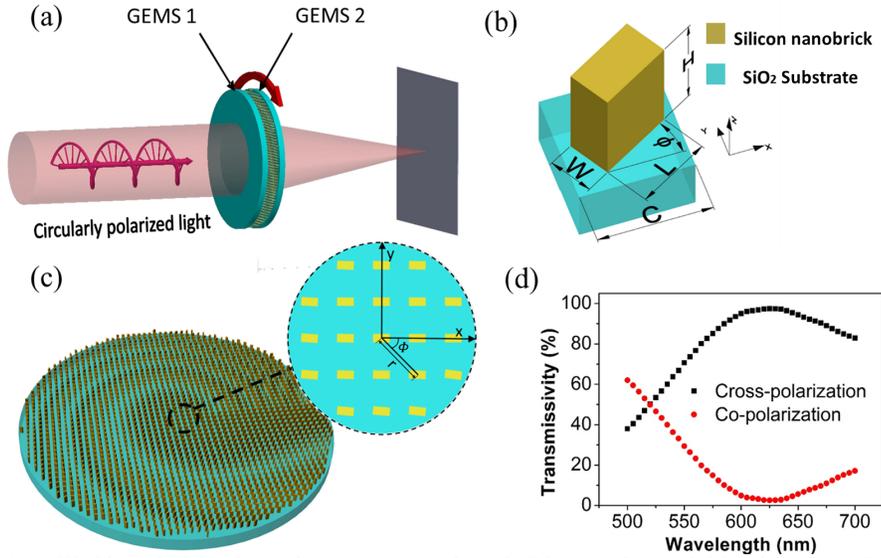

Fig. 1. (a) Working principle of a continuous-zoom metalens. (b) Diagram of a nanobrick unit structure. Each nanobrick was designed with length $L$ = 140 nm, width $W$ = 70 nm, height $H$ = 350 nm and cell size $C$ = 300 nm at an operation wavelength of 633 nm. $\phi$ is the orientation angle of the nanobrick and the incident light propagates along $z$-axis direction. (c) Schematic diagram of one of the designed GEMSs. (d) The co-polarization and cross-polarization transmissivities of a nanobrick. The simulation was run by normally illuminating a nanobrick with a beam of circularly polarized light and under the boundary condition of period.

The geometry of a nanobrick was designed and simulated by the commercial electromagnetic simulation software (Comsol). To simulate the effect of a single nanobrick in an array, periodic boundary condition was used in the modeling process. The nanobrick was illuminated by CP light, we swept the nanobrick's dimension and cell size at an operation wavelength of 633 nm to optimize the performance. As shown in Fig. 1(d), the transmitted light consists of both circular polarization states: the cross-polarization with a phase delay $\varphi$ is the opposite handedness to that of the incident CP light, while the co-polarization has the same handedness without phase delay. This configuration with optimized dimension maintains a fairly large transmissivity of cross-polarization over 95% and the undesired co-polarization is extremely low which means that each nanobrick as a sub-wavelength half wave plate lead to nearly complete conversion between two opposite circular polarization states.

For a Moiré diffractive lens, the transmission functions of two phase plates can be described as [26, 36, 37]

$$T_1(r, \Phi) = \exp[iar^2\Phi] \tag{1}$$

and

$$T_2(r, \Phi) = \exp[-iar^2\Phi], \tag{2}$$

where $r$ is the radius from an arbitrary point to the center of the phase plate, $\Phi$ is the angle between $x$

axis and the line where the point is located to the center as shown in Fig. 1(c), and *a* is a constant. If we use GEMSs to form the two phase plates, due to the polarization dependence of a GEMS, the left-handed circularly polarized (LCP) incident light will be converted into right-handed circularly polarized (RCP) light after passing through the first GEMS, that is, the second GEMS is illuminated by a beam of RCP light. Since the two equations are complex conjugates and the signs of the phase change amount generated by LCP and RCP states are reversed, it is interesting that the two GEMSs can be designed with the same orientation angle distribution of nanobricks. If we combine two GEMSs face to face as shown in Fig. 1(a), the two nanobrick arrays should be chiral.

When the second GEMS is rotated by a mutual angle $\theta$ around its optical axis, the transmission function $T_2$ is converted to

$$T_{2,rot}(r,\Phi) = \exp[-iar^2(\Phi-\theta)]. \qquad (3)$$

Therefore, the transmission function of the metalens is

$$T_{com} = T_1(r,\Phi)T_{2,rot}(r,\Phi) = \exp[ia\theta r^2]. \qquad (4)$$

Compared with the transmission function of a spherical lens, namely,

$$T_{lens} = \exp[i\frac{\pi r^2}{\lambda f}], \qquad (5)$$

where $\lambda$ is the operation wavelength and *f* is the focal length of a lens, the focal length of the metalens satisfies

$$f = \frac{\pi}{a\theta\lambda} \qquad \text{for LCP incident light}, \qquad (6)$$

and

$$f = -\frac{\pi}{a\theta\lambda} \qquad \text{for RCP incident light}. \qquad (7)$$

From Eqs. (6) and (7), we can observe that when the constant *a* and the incident wavelength $\lambda$ are determined, the focal length will only depend on the mutual rotation angle $\theta$, in other words zoom functionality of the continuous-zoom metalens can be achieved by mutually rotating two GEMSs. The focal length vs rotation angle is shown in Fig. 2. Due to the polarization dependence of a GEMS, if the metalens is designed as a positive lens in LCP light incidence, it will become a negative lens when we change the polarization state of the incident light, and vice versa. What the most important thing is that it can make the focal length of the metalens transform instantaneously from a positive value to a negative one whilst retains the same absolute value by changing the handedness of the CP incident light. However, because of the periodicity of mutual rotation, there is a discontinuous spot at 0 degrees or say 360 degrees and the result at this angle can only be described as no focusing or focusing on infinity.

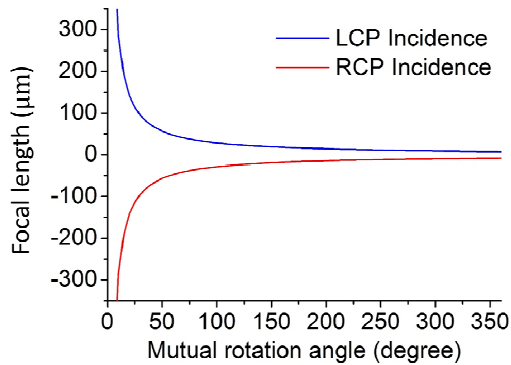

Fig. 2. The focal length of the metalens as a function of the mutual rotation angle between two GEMSs. With LCP incident light, the metalens acts as a positive zoom lens and its focal length changing from $+\infty$ to nearly zero like the blue curve. When the incident light is switched to RCP, the metalens becomes a negative one and its focal length ranges from $-\infty$ to nearly zero. In this example, the metalens dimension is 30.3 × 30.3 μm², the constant *a* is 0.1 μm⁻² and the operation wavelength is 633 nm.

In Eqs. (6) and (7), the constant $a$ which is limited by the resolution of a GEMS depends on the pixel size and the maximum radius of the GEMS, and its constraint can be expressed by [26]

$$a < \frac{1}{2Cr_{max}}, \qquad (8)$$

where $C$ is the cell size of a nanobrick structure and $r_{max}$ is the radius of the GEMS.

From Eqs. (6) and (7), we can find that the minimum value of the focal length $f$ is related to the cell size $C$ and the GEMSs' radius $r_{max}$. For example, with a pixel size of $300 \times 300$ nm$^2$, lens dimension of $3.3 \times 3.3$ μm$^2$ ($11 \times 11$ pixel numbers) and operation wavelength of 633 nm, the minimum focal length is calculated to be a value of only 780 nm when $\theta$ is near to 360°. This value (780 nm) is even close to the wavelength and we can consider that it is near to zero in this working condition. However, because of the periodicity shown in Eq. (4), in addition to the desired sector, there is another sector which takes up a part of incident light power appears, and the scale of the undesired sector is proportional to $\theta$ [26]. As a consequence, the energy falling in the focal spot decreases as the mutual rotation angle $\theta$ keeps increasing. In order to ensure that the metalens satisfies the request of enough focal spot energy, we set the maximum angle of $\theta$ to 180° and the value of minimum focal length becomes 1.56 μm which is still an extremely short length close to twice as long as the operation wavelength. On the other hand, if the metalens is designed by Eq. (6) with a positive focal length, we can easily observe that the focal length can be extended to $+\infty$ when $\theta$ is near to zero. Therefore, by rotating one GEMS, the metalens can cover a focal length ranging from $+\infty$ to an extremely short length. More importantly, only by changing the handedness of CP incident light, the focal length can cover nearly $-\infty \sim 0$. Finally, we can obtain a continuous-zoom metalens which contains a varing focal length almost from $-\infty$ to $+\infty$ (except for a relatively short gap) only by rotating any one of the GEMSs along its optical axis.

## 3. Numerical simulations and discussion

### 3.1 Numerical simulations of a zoom metalens

For practice, we designed a continuous-zoom metalens and simulated the light propagation with a beam of normally incident CP light by using the commercial electromagnetic simulation software package (FDTD Solutions). Three-dimensional finite-difference time-domain method was used in the simulations and the boundary condition was perfect matching layers (PML). The two GEMSs of the simulated metalens with phase distribution of $\Phi = a\theta r^2$ have a cell size of $300 \times 300$ nm$^2$, aperture dimension of $30.3 \times 30.3$ μm$^2$ and a constant $a$ of 0.1 μm$^{-2}$, operating at a wavelength of 633 nm. The simulation results are shown in Fig.3. Due to the limitation of computation resource and time, we can only simulate a small aperture metalens at several selected rotation angles.

The fact is that it is most ideal state when two phase profiles of GEMSs overlay completely. However, for the purpose of continuous zooming by rotating, the two GEMSs are separated by a distance of 140 nm which brings about that the transmitted light deviates lightly from the optical axis, such out of symmetry also stems from that the phase distribution of each GEMS is not centrally symmetric. The intensity distribution of the electric field in three situations ($\theta$ is 30°, 45° and 75°, respectively) indicates excellent light focusing without noise spots, shown in Figs. 3(a-c). It is observed that the focal planes of $\theta = 30°$, 45° and 75° in Figs. 3(a-c) have different deviations from 94.8 μm, 63.2 μm and 37.9 μm calculated theoretically by Eq. (6). It is because of the coupling effect among adjacent nanobricks and finite aperture of the metalens [38, 39], such deviation will decrease as the mutual rotation angle $\theta$ increases or say the focal length decreases. Although the simulated focal plane is not at an ideal place, it does not affect the zoom functionality of a large range since the mutual rotation angle $\theta$ is reconfigurable and all the simulation results show that the metalens can focus light well. In a practical application, we can calibrate the exact focal length for different mutual rotation angles. In addition, the calculated focusing efficiencies (i.e. the ratio of focal spot energy at focal plane and incident energy) of the metalens are 35.92%, 32.92% and 26.67% corresponding to $\theta = 30°$, 45° and 75°, respectively, which agree well with the correlative relationship between focal spot energy and the mutual rotation angle of the metalens.

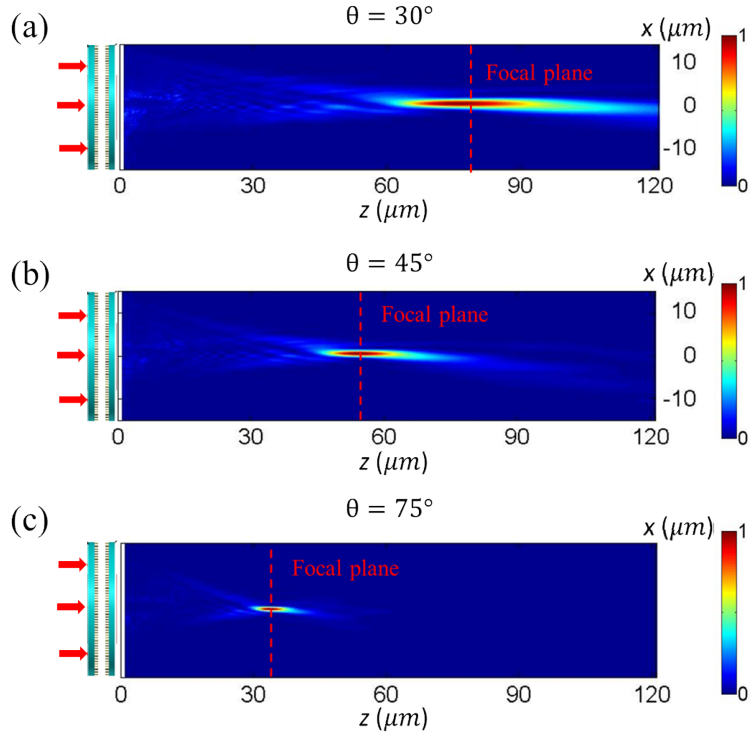

Fig. 3. Simulation of a zoom metalens. Full-wave numerical simulations were performed by FDTD Solutions for the propagation of LCP incident light. (a, b, c) Intensity of electric field along the direction of light propagation (z axis) corresponding to $\theta = 30°$, 45° and 75°, respectively. The back surface of metalens is placed at z = 0 μm.

The resolution of a metalens is determined by the diffraction limit with

$$\sigma = \frac{0.61\lambda}{NA}, \qquad (9)$$

where *NA* is the numerical aperture of the metalens. On the basis of Eq. (9), a metalens with a longer focal length corresponds to a smaller *NA* and would produce a larger spot for the same lens aperture. In Figs. 4(a-c), for the same aperture (30.3 μm), the focal spot size at the focal plane of the optical system varies with different focal lengths, which keeps consistent with the diffraction limit principle.

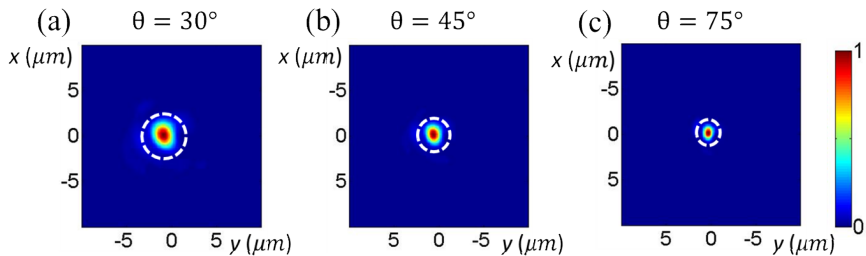

Fig. 4. Simulation results at the focal plane of the metalens. (a, b, c) Intensity of electric field at cross section of focal plane corresponding to $\theta = 30°$, 45° and 75°, respectively. The airy disks marked by white curves are calculated with the diffraction limit principle and their radii are 2.42 μm ($\theta = 30°$), 1.61 μm ($\theta = 45°$) and 0.97 μm ($\theta = 75°$), respectively.

The simulation results in Fig. 5 show that the metalens transforms from a positive lens to a negative one when the polarization state of incident light is changed from LCP to RCP, i.e., the focal length of θ = 75° changes from 37.9 μm to −37.9 μm in Fig. 5(a) and the focal length 94.8 μm of θ = 30° becomes −94.8 μm in Fig. 5(b). The simulation results reveal that incident light passes through the metalens does not focus anymore and the curvature radii of wave fronts corresponding to 75° and 30° are approximately 41.2 μm and 109.6 μm, respectively. It turns out that the output light after the metalens is divergent in image space and its negative focal length also can be adjusted continuously by changing the mutual rotation angle. In contrast to the negative focusing, Figs. 5(c-d) show the situations of positive focusing when θ is 75° and 30°. This confirms that the focal length of the metalens described in this paper can cover a fairly large dynamic range by rotating two chiral GEMSs of a metalens and changing the polarization state of incident light.

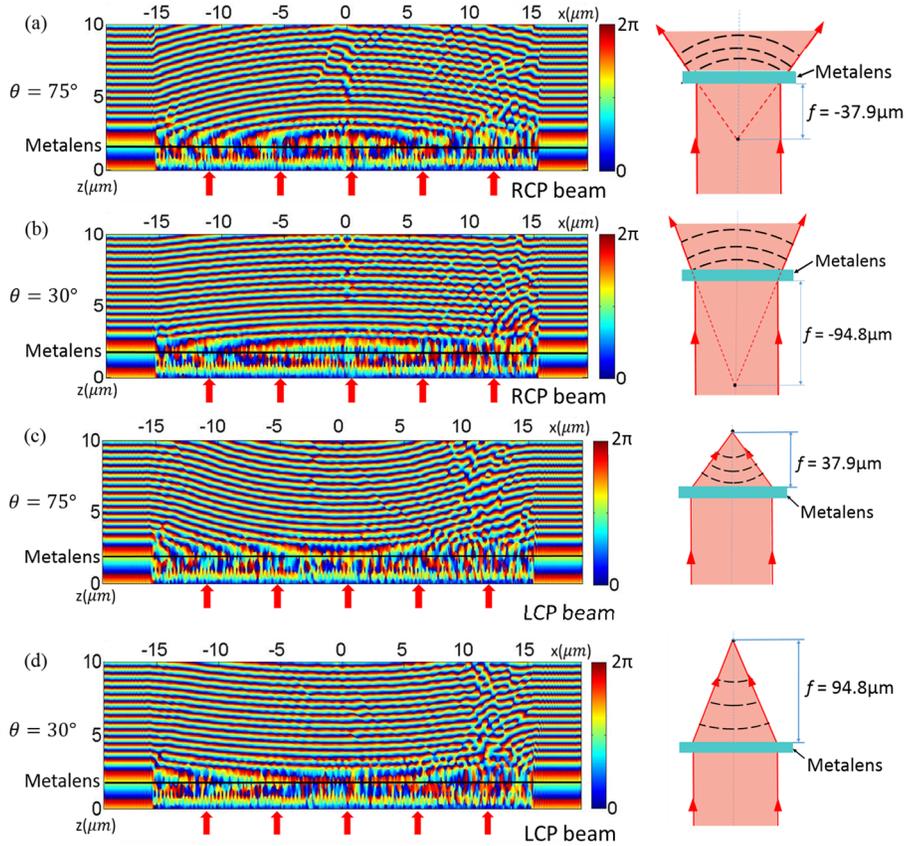

Fig. 5. Simulation of a zoom metalens. (a, b) The simulation domains show the distribution of phase along the direction of light propagation (z axis) for θ = 75° (f = −37.9 μm) and θ = 30° (f = −94.8 μm), respectively. (c, d) Comparing with the negative focal length case, the simulation domains show the phase distribution of θ = 75° (f = 37.9 μm) and θ = 30° (f = 94.8 μm) with LCP light incidence.

## 3.2 Numerical simulations of a corrected metalens with a large zoom range

It is known that the metalens obtains the maximum focusing energy at θ = 0°; however, the focal length is infinity which does not satisfy practical requirement. The contradiction between the focusing energy and the focal range can be solved by adding a constant phase compensation [36] and the transmission functions become

$$T_1(r, \Phi) = \exp[iar^2\Phi + ibr^2] \tag{10}$$

and

$$T_2(r,\Phi) = \exp[-iar^2\Phi + ibr^2], \qquad (11)$$

where $b$ is a constant. Therefore, a new combination metalens with a mutual rotation $\theta$ can be described as

$$T_{com}(r,\Phi) = \exp[-iar^2\theta + i2br^2]. \qquad (12)$$

The focal length of the corrected metalens becomes

$$f = \frac{\pi}{(a\theta + 2b)\lambda} \qquad \text{for LCP incident light,} \qquad (13)$$

and

$$f = -\frac{\pi}{(a\theta + 2b)\lambda} \qquad \text{for RCP incident light.} \qquad (14)$$

In Eqs. (13) and (14), it demonstrates that the focal length reaches its maximum at $\theta_{min} = -2b/a$, and the constant $b$ depends on the selection of $\theta_{min}$. On the other hand, the maximum diffraction efficiency is still generated at $\theta = 0°$.

To further verify zoom functionality of the corrected metalens with a phase-compensation, we assume that the mutual rotation angle $\theta$ increases from $-90°$, that is, the focal length reaches a maximum at $-90°$, which means that the constant $b$ in Eqs. (13) and (14) equals to $a\pi/4$. To show the overall range variation of the focal length vs $\theta$, we selected four certain mutual rotation angles $\theta$ which equal $-90°$, $-45°$, $0°$ and $45°$, respectively. The four simulation results in Fig. 6 show that the focal length of the corrected metalens changes from $-\infty$ to a relative short one. By means of introducing phase compensation, the incident light is focused at infinity when $\theta = 90°$ which is different from the original metalens that focuses light to infinity at $\theta = 0°$. Instead, the focal length is reduced to 31.5 µm at $\theta = 0°$. For the focal spots at three different focal planes of $\theta = -45°$, $0°$ and $45°$, the calculated focusing efficiencies are 27.37%, 37.68% and 26.61%, respectively. Furthermore, the corrected metalens still has the functionality of polarization control, namely, as a beam of incident light converts its polarization state into an opposite one, the polarity of the metalen also switches rapidly between positive one and negative one. We assume that the positive and negative focusing efficiencies are both less than our requirement when $\theta$ is greater than 180°, then the zoom range of the corrected metalens is broader than that of the original one within the required efficiency.

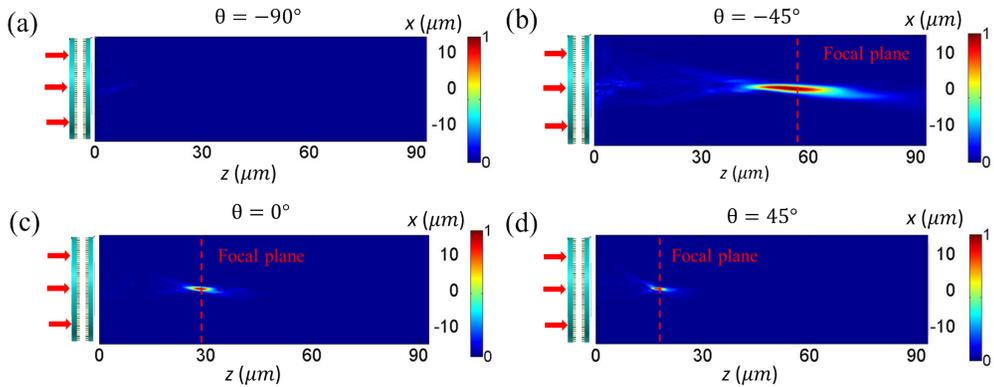

Fig.6. A variation tendency of the focal length for the corrected metalens. (a, b, c, d) Intensity of electric field along the direction of light propagation (z axis) corresponding to $\theta = -90°$ ($f = +\infty$), $\theta = -45°$ ($f = 63.1$ µm), $\theta = 0°$ ($f = 31.5$ µm) and $\theta = 45°$ ($f = 21.0$ µm), respectively. The corrected metalens is illuminated by a beam of

LCP incident light.

### 3.3 Discussion

The proposed continuous-zoom metalens designed with two GEMSs have great advantages over the previously proposed zoom diffractive lenses. Firstly, the pixel size of GEMS is at a sub-wavelength level so that the phase profile of a metalens is much finer. In this paper, the total size of the simulated metalens is 101 × 101 pixels and each pixel size is 300 × 300 nm$^2$, which is smaller than half of a wavelength and can provide a relatively high resolution and information density. Secondly, the metalens based on GEMSs can manipulate phase continuously which makes the phase distribution more continuous and the process of zooming smoother. Compared with most of existing continuous-zoom metalenses, the metalens we proposed does not require any spatial extension like lateral or axial displacement, which provides a better implementation for ultrathin and reliable imaging systems. Finally, because of the polarization dependence of a GEMS, the positive and negative polarities are interchangeable in one identical metalens only by changing the handedness of incident circularly polarized light. Therefore, we can manipulate the focal length of a metalens in a large dynamic range while not at the expense of increased complexity of an imaging system. In addition, such a wide zoom range has never been implemented by previously proposed metalens.

In applications, the high integration and simple structure of the reconfigurable continuous-zoom metalens makes it promising in consumer electronics. It can be used as an integrated camera in mobile phones and smart devices. It can also be integrated into a three-dimensional (3D) light sensing equipment to flexibly adjust the imaging range, making it easier to be used in a 3D sensing application. The above zoom technology can be further extended, such as a microwave scanning antenna, which can realize continuous and flexible switching of the microwave in different fields of view.

### 4. Conclusions

In summary, we propose a continuous zoom metalens, which is composed of only two metasurface-based phase plates with precise phase modulation and ultracompact structures. The zooming process is realized by mutually rotating two GEMSs, which is simple in operation and fast in response. Besides, the zoom metalens can provide fairly high information density, finer phase distribution and the focal length can almost vary from −∞ to +∞ without spatial extension in any direction or additional equipment. Compared with conventional DOE-based zoom lens, the metasurfaces are two-dimensional (2D) planar material so it doesn't require the multi-step design and process. With such a miniaturized structure and simple design, the zoom metalens can be applied to many optical imaging systems with ultra-compact and continuous-zoom requirements.

### Acknowledgements

National Natural Science Foundation of China (Numbers 11774273, 11574240, 61640409 and 61805184); Outstanding Youth Funds of Hubei Province (Number 2016CFA034); Open Foundation of State Key Laboratory of Optical Communication Technologies and Networks, Wuhan Research Institute of Posts and Telecommunications (Number OCTN-201605); Postdoctoral Innovation Talent Support Program of China (BX20180221); Guangxi Natural Science Foundation of China (Number 2017GXNSFAA198048).